\documentclass[reprint,amsmath,amssymb,aps]{revtex4-2}

\usepackage{graphicx}% Include figure files
\usepackage{color}
\usepackage{dcolumn}% Align table columns on decimal point
\usepackage{bm}% bold math
\usepackage{nicefrac}% add hypertext capabilities
\usepackage{hyperref}% add hypertext capabilities
\usepackage[textsize=tiny]{todonotes}

\renewcommand{\eqref}[1]{Eq.~(\ref{#1})}
\newcommand{\figref}[1]{Fig.~\ref{#1}}
\newcommand{\secref}[1]{Sec.~\ref{#1}}

\newcommand{\refcite}[1]{Ref.~\cite{#1}}
\newcommand{\refscite}[1]{Refs.~\cite{#1}}

\begin{document}

\title{Theory and simulation of multiphase coexistence in biomolecular mixtures}

\author{William M.~Jacobs}
\email{wjacobs@princeton.edu}
\affiliation{Department of Chemistry, Princeton University, Princeton, NJ 08544, USA}

\date{\today}

\begin{abstract}
  Biomolecular condensates constitute a newly recognized form of spatial organization in living cells.
  Although many condensates are believed to form as a result of phase separation, the physicochemical properties that determine the phase behavior of heterogeneous biomolecular mixtures are only beginning to be explored.
  Theory and simulation provide invaluable tools for probing the relationship between molecular determinants, such as protein and RNA sequences, and the emergence of phase-separated condensates in such complex environments.
  This review covers recent advances in the prediction and computational design of biomolecular mixtures that phase-separate into many coexisting phases.
  First, we review efforts to understand the phase behavior of mixtures with hundreds or thousands of species using theoretical models and statistical approaches.
  We then describe progress in developing analytical theories and coarse-grained simulation models to predict multiphase condensates with the molecular detail required to make contact with biophysical experiments.
  We conclude by summarizing the challenges ahead for modeling the inhomogeneous spatial organization of biomolecular mixtures in living cells.
\end{abstract}

\maketitle

\section{Introduction}

The discovery that intracellular ``organelles'' can exist without membranes has revolutionized molecular and cellular biology~\cite{brangwynne2009germline,shin2017liquid}.
Many such intracellular structures, now collectively referred to as ``biomolecular condensates,'' have been proposed to form via phase separation~\cite{shin2017liquid,berry2018physical,choi2020physical,villegas2022molecular}.
Physically, this means that a surface tension holds the phase-separated condensate together, while individual biomolecules---including proteins, RNAs, and other small molecules---exchange between the condensate and the surrounding fluid in dynamic equilibrium.
Phase-separated condensates represent a unique form of biological organization compared to traditional membrane-bound organelles, since the absence of a membrane allows for rapid assembly and disassembly in response to stimuli~\cite{berry2018physical}.

Over the past 15 years, an increasingly large number of biomolecular condensates have been identified~\cite{banani2017biomolecular}.
Because of the wide range of biological phenomena in which condensates play a role, including both fundamental biological processes~\cite{boija2018transcription,sabari2018coactivator,wang2018single,kim2019phospho,lafontaine2021nucleolus,klosin2020phase,su2016phase,tiwary2019protein} and a variety of pathological conditions~\cite{alberti2021biomolecular,jack2021sars}, it is important to understand the biophysical mechanisms that control which biomolecules partition into specific condensates.
Theoretical advances are needed to guide experiments probing the relationship between the properties of individual biomolecules and emergent condensate structures in complex environments.
In particular, the physicochemical determinants of condensate composition and stability in heterogeneous intracellular environments---where thousands of biomolecular species are present---are only beginning to be explored.
This review summarizes theoretical and simulation efforts in this direction using approaches based on equilibrium thermodynamics.

\subsection{Linking physicochemical properties and condensate thermodynamics}

\begin{figure}
  \includegraphics[width=8.5cm]{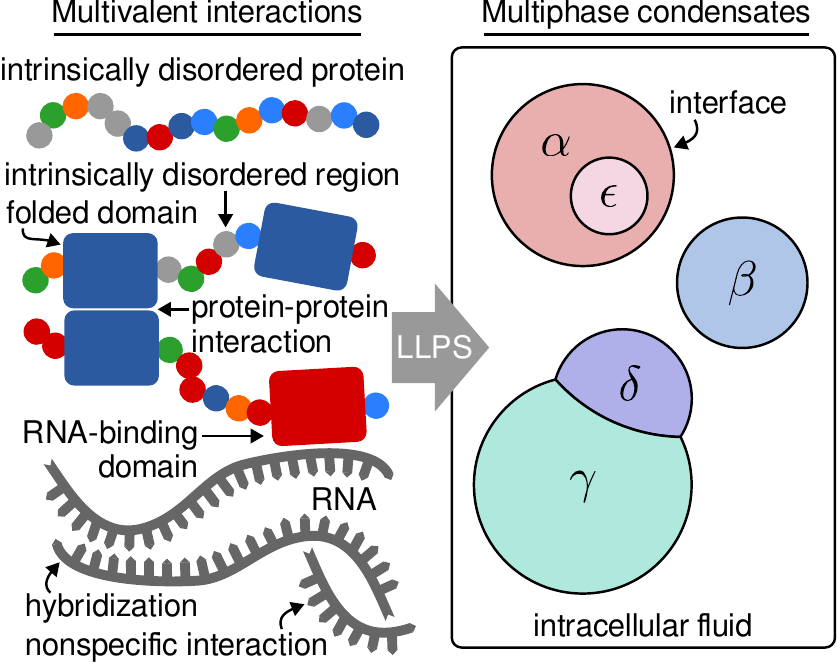}
  \caption{Multivalent interactions among a wide variety of biological macromolecules, including intrinsically disordered proteins (with amino acids represented by colored circles), multidomain proteins, and nucleic acids, contribute to the thermodynamic driving forces responsible for liquid--liquid phase separation.  Phase-separated condensates, including higher-order structures composed of multiple immiscible phases, resemble ``membraneless organelles'' whose interfaces are stabilized by surface tensions.  The molecular compositions within each phase ($\alpha$--$\epsilon$) are distinct as a result of specific interactions among the constituent biomolecules.}
  \label{fig:biomolecules}
\end{figure}

How do biomolecular determinants such as amino-acid or nucleotide primary sequence, secondary/tertiary structure, and chemical modifications control the compositions and spatial organization of phase-separated intracellular condensates (\figref{fig:biomolecules})?
This question has been addressed primarily within the context of equilibrium thermodynamics, in which the phase behavior of a macromolecular mixture is governed by free energies at thermal equilibrium.
Within this framework, the partitioning of biomolecules into phase-separated condensates is determined by equilibrium chemical potentials, while condensate (dis)assembly dynamics are governed by free-energy gradients close to equilibrium and/or transitions between metastable states.
Predictions based on this near-equilibrium assumption generally hold up well when tested against \textit{in vitro} experiments~\cite{berry2018physical,choi2020physical,wei2020nucleated}.
Thus, while living systems may be more accurately characterized as nonequilibrium steady states under some conditions~\cite{soding2020mechanisms}, we will restrict our attention to near-equilibrium approaches for predicting biomolecular phase separation in this review.
We will also use the common terminology \textit{liquid--liquid phase separation (LLPS)}~\cite{colby2003polymer,mcmaster1975aspects,taratuta1990liquid,hyman2014liquid,shin2017liquid} to describe reversible thermodynamic phase transitions between (potentially complex) fluid phases with different macromolecular concentrations, as our discussion will focus on static properties such as condensate composition and spatial organization.
Nonetheless, we note that condensed phases in biology often exhibit viscoelastic dynamical properties and may irreversibly age into solid phases due to the complexity of the interactions among biological macromolecules~\cite{banani2017biomolecular,hyman2014liquid,jawerth2020protein,mittag2022conceptual,alshareedah2022determinants,tejedor2022protein,alshareedah2023sequence}.

Concepts from polymer physics have helped shape the prevailing view that transient associations among biomolecules give rise to the overall net attractive interactions required to bring about LLPS~\cite{brangwynne2015polymer}.
These interactions are commonly referred to as ``multivalent,'' since biomolecules can associate through multiple interaction sites via a variety of forms of noncovalent bonding.
Particular attention has been given to conformationally heterogeneous proteins, including intrinsically disordered proteins (IDPs) and multidomain proteins containing intrinsically disordered regions (IDRs)~\cite{brangwynne2015polymer}.
In the context of IDPs, multivalency refers to the ability of an unfolded protein to engage in many residue--residue contacts with nearby proteins in a condensed phase.
Folded domains within multidomain proteins can also contribute to the multivalency required to drive LLPS, either through protein--protein interactions (PPIs)~\cite{sanders2020competing} or, in the case of RNA binding domains (RBDs), through interactions with RNA~\cite{lin2015formation}.
Finally, nucleic acid mixtures can phase separate under certain conditions due to intermolecular base-pairing~\cite{king2021phase,rovigatti2014accurate,hegde2023competition} and nonspecific association~\cite{wadsworth2022rnas}.
Importantly, the strengths of the net interactions among biopolymers in liquid-like condensates are typically comparable to the thermal energy, since the protein and nucleic acid constituents of biomolecular condensates can often remain fluid on biologically relevant timescales.

\subsection{Emergence of multiphase coexistence in complex biomolecular mixtures}

Biological LLPS results in an enormous diversity of condensates in living cells.
Each of these condensates is associated with a specific chemical composition~\cite{ditlev2018who} and may be enriched in many distinct biomolecules relative to the surrounding intracellular fluid~\cite{xing2020quantitative}.
The biological functions of condensates derive directly from this compositional specificity, since the biochemical reactions that take place within the spatial confines of a condensate are dependent on the molecular concentrations that define the local environment.
Theoretical descriptions of \textit{in vivo} condensate assembly must therefore account for complex intracellular mixtures comprising thousands of protein and RNA species, which can all potentially interact with one another.

At the simplest level, it is important to distinguish between homotypic and heterotypic interactions between species of the same or different types, respectively.
In multicomponent mixtures with strong heterotypic interactions, the tendency of any particular species to partition into a condensate depends on the concentrations of all its potential interaction partners~\cite{riback2020composition}.
A consequence is that the equilibrium compositions of coexisting phases may depend on the concentrations of all the components in the mixture, even when there are only two phases in coexistence.
This feature can be used to detect the influence of multiple components on phase separation and to infer the relative strengths of homotypic and heterotypic interactions by measuring the volume fractions of coexisting phases at different overall mixture concentrations~\cite{qian2022tie,hegde2023competition}.

Multiple immiscible condensates are commonly found to coexist within a single intracellular compartment~\cite{banani2017biomolecular}.
Moreover, depending on the properties of the interfaces between pairs of condensates and between condensates and the surrounding fluid, immiscible condensates can self-organize into spatially organized structures~\cite{fare2021higher}.
Well characterized examples include the nucleolus~\cite{feric2016coexisting,lafontaine2021nucleolus} and stress-granule/P-body condensates~\cite{sanders2020competing,guillen2020rna,yang2020g3bp1}.
It has also become clear that subtle changes in protein and RNA concentrations can perturb the interfacial properties and thus dramatically alter the architecture of multiphasic condensates~\cite{sanders2020competing,kaur2021sequence}.
Nonetheless, predicting multiphase coexistence in the context of heterogeneous intracellular fluids remains a formidable challenge.

\subsection{Aims and scope of this review}

\begin{figure}
  \includegraphics[width=8.5cm]{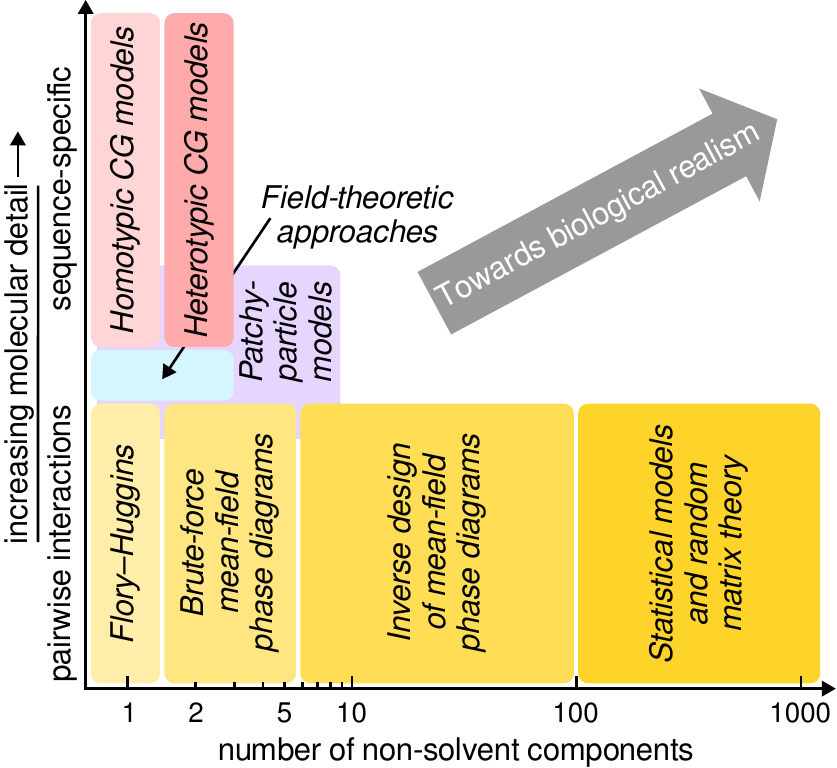}
  \caption{Computational and theoretical complexity increases with both the level of molecular detail and the number of distinct components in a mixture.  Simulation approaches to biomolecular LLPS range from pairwise-interaction mean-field models to sequence-specific coarse-grained (CG) models.  However, mixtures with more than three non-solvent components have so far been studied almost exclusively using pairwise mean-field models.}
  \label{fig:roadmap}
\end{figure}

Developing theoretical and computational models of multiphasic, multicomponent biomolecular mixtures is essential for understanding the relationship between molecular determinants and biological self-organization via LLPS.
The purpose of this article is to highlight a number of advances in this direction.
Many recent reviews focusing on theory and simulation, including \refscite{choi2020physical}, \cite{dignon2020biomolecular}, and \cite{lin2018theories}, have described coarse-grained modeling approaches for IDPs, multidomain proteins, and nucleic acids.
These approaches have primarily been applied to study the properties of single molecules and to mimic \textit{in vitro} experiments on condensate formation.
By contrast, we focus here on theoretical challenges that arise when considering multiphase coexistence, especially in mixtures with thousands of components.
Studies along these lines have provided complementary insights that are needed to understand biomolecular condensates in an intracellular context (\figref{fig:roadmap}).
For broader context, we encourage the reader to consult other recent works, including reviews that emphasize the biological functionality and regulation of condensates~\cite{lyon2021framework,alberti2021biomolecular}, the interplay between physical gelation and phase separation of multivalent macromolecules~\cite{pappu2023phase}, and the conformational dynamics of macromolecules within condensates~\cite{abyzov2022conformational}.

In this review, we begin in \secref{sec:thermodynamics} by covering the thermodynamic principles of phase separation in multicomponent fluids.
We highlight recently devised numerical methods for calculating multiphase coexistence in both mean-field and classical molecular simulation models.
We then discuss theoretical results obtained from mean-field multicomponent mixture models in \secref{sec:mean-field}.
These studies have provided important insights into phase-behavior scaling relations, although they lack molecular detail and, as such, require assumptions on the statistical properties of intermolecular interactions in complex fluids.
In \secref{sec:sequence-dependent}, we examine efforts to describe multicomponent condensates with both analytical and computational models that capture the molecular sequence dependence or the structure of a PPI network.
The implications of these studies for the mean-field multicomponent mixture models introduced in \secref{sec:mean-field}, and potential extensions thereof, are discussed.
Finally, in \secref{sec:outlook}, we identify key challenges that must be overcome in order to describe inhomogeneous spatial organization in living cells with molecular realism.

\section{Thermodynamic principles of multicomponent LLPS}
\label{sec:thermodynamics}

\begin{figure}
  \includegraphics[width=8.5cm]{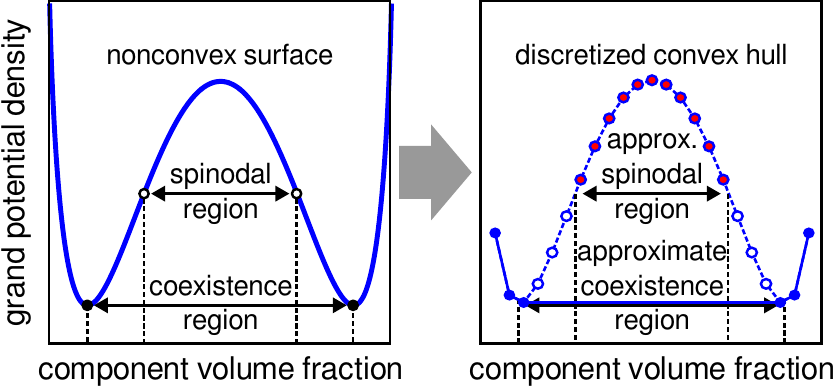}
  \caption{\textit{Left:} Nonconvex free-energy surfaces lead to phase separation at thermodynamic equilibrium.  The inflection points and global minima of the grand potential density, ${\Omega \equiv f - \sum_{i=1}^N \rho_i \mu_i}$, determine the spinodal points and coexistence points, respectively.  \textit{Right:} Approximate phase diagrams can be obtained by computing the convex hull (solid line) of a discretized free-energy surface; points on the hull (filled circles) are in one-phase regions, while points not on the hull (empty and red-filled circles) are within a coexistence region.  The approximate spinodal region can be determined by identifying points where the Hessian is not positive definite (red-filled circles).  Approximate coexistence points can then be refined via nonlinear minimization (see text).  This scheme generalizes to higher-dimensional concentration spaces.}
  \label{fig:phase-diagram}
\end{figure}

Phase coexistence describes an equilibrium state in which a material or fluid exists in multiple phases with distinct physicochemical properties, such as oil droplets suspended in aqueous solution.
Thermodynamic equilibrium between coexisting phases is established when the temperature, (osmotic) pressure, and chemical potentials of all molecular species are constant throughout the system.
Considering a biomolecular solution at constant volume and temperature, the thermodynamic state of the system can be described by the Helmholtz free-energy density, $f$.
This free energy is a function of the concentrations, $\{\rho_i\}$, of all the molecular components in the mixture.
(Latin indices will be used throughout to indicate molecular components, while Greek indices will be used to indicate phases.  Analogous arguments apply to the Gibbs free-energy density in the case of fluids at constant pressure.)
Phase separation can occur when the free-energy density is a nonconvex function of the molecular concentrations (\figref{fig:phase-diagram}).
In such a case, the free energy can be minimized by forming two or more distinct phases---for example, a condensed droplet and the surrounding cytoplasm---each with different concentrations.
A mixture phase separates when the overall concentrations of the solution lie within the \textit{coexistence region}, which is bounded by the concentrations of the coexisting phases.
Droplets that emerge as a result of this spontaneous process are stabilized by positive surface tensions at the interfaces that form between the coexisting phases.
Whenever $f$ is nonconvex, there is also a \textit{spinodal region} within which the free-energy surface has negative curvature.

In a heterogeneous system comprising many different types of biomolecules, the free-energy surface is a high-dimensional object.
Nonetheless, coexistence and spinodal regions can still be determined by examining the convexity and local curvature of the free-energy surface.
More precisely, the Hessian matrix $\partial^2 f/\partial\rho_i\partial\rho_j$ is not positive definite within the spinodal region, implying that a homogeneous mixture within this region is unstable with respect to concentration fluctuations in one or more directions of concentration space.
These directions are described by the eigenvectors that correspond to the negative eigenvalues of $\partial^2 f/\partial\rho_i\partial\rho_j$.
The region of a high-dimensional concentration space in which concentration fluctuations are locally unstable is bounded by a spinodal locus, where the determinant $|\partial^2 f/\partial\rho_i\partial\rho_j|=0$.

The molecular concentrations of coexisting bulk phases can be determined by considering the equal pressure and chemical potential conditions.
In multicomponent fluids, these conditions can be satisfied by performing a ``common tangent plane construction,'' in which a hyperplane is tangent to the free-energy surface at each point in concentration space that corresponds to a coexisting stable phase.
A homogeneous mixture with an overall, or ``parent'', concentration vector inside the convex hull of the coexisting-phase concentrations can lower its Helmholtz free energy by phase-separating.
This convex hull therefore defines the coexistence region, which necessarily encompasses the spinodal region, in a multicomponent fluid.
Because the tangent at any point on the free-energy surface is equal to the chemical potential vector, $\{\mu_i\} = \partial f/\partial \rho_i$, the common tangent plane construction ensures equal chemical potentials for each species across all phases that are in coexistence.
Furthermore, the common tangent plane construction implies that the coexisting phases are all global minima of the grand potential density, $\Omega(\{\rho_i\};\{\mu_i\}) \equiv f(\{\rho_i\}) - \sum_i \rho_i \mu_i$.
This fact ensures equal pressures among all bulk phases.

In general, the coexistence concentrations in a multicomponent fluid are not specified uniquely without also prescribing the parent concentrations, $\{\rho_i\}^{\text{(parent)}}$.
The connection between the parent and coexisting-phase concentrations is provided by the conservation law
\begin{equation}
  \label{eq:lever-rule}
  \rho_i^{(\text{parent})} = \sum_{\alpha=0}^{K} x^{(\alpha)} \rho_i^{(\alpha)}(\{\mu_j\})\quad\forall i,
\end{equation}
where $\alpha$ indexes the phases in a phase-separated state with $K+1$ phases, the concentrations $\{\rho_i^{(\alpha)}\}$ indicate coexisting phases with coexistence chemical potentials $\{\mu_j\}$, the volume fractions of the bulk phases are given by $\{x^{(\alpha)}\}$, and $\sum_{\alpha=0}^K x^{(\alpha)} = 1$.
(This indexing convention is chosen for later convenience, since we are often interested in phase equilibria involving a solvent-majority phase, $\alpha = 0$.)
\eqref{eq:lever-rule} simplifies to the well-known lever rule for binary mixtures (e.g., fluids comprising one macromolecular component plus a solvent).

The spinodal locus coincides with the boundary of the coexistence region at a critical point, where the concentrations of two coexisting phases merge into a single stable phase.
Unlike binary mixtures, there is typically no unique critical point in a multicomponent fluid.
Instead, multicomponent critical points lie on a temperature-and-concentration-dependent manifold with dimension one less than the number of non-solvent components.
Higher-order critical points, where more than two phases simultaneously merge into a single stable phase, are also possible in multicomponent fluids~\cite{griffiths1970critical}.

Multicomponent phase equilibria can equivalently be determined from the excess chemical potential, $\mu_{\text{ex},i}$, of each molecular species $i$.
This quantity represents the contribution to the chemical potential that captures all interactions---both enthalpic and entropic---among the molecules, and is thus a function of all the component concentrations~\cite{chandler1987introduction}.
The excess chemical potential is directly related to the partition coefficient, PC, defined as the ratio of a molecule's concentration inside (in) and outside (out) of a phase-separated droplet:
\begin{equation}
  \label{eq:PC}
  \text{PC}_i \equiv \frac{\rho_i^{\text{(in)}}}{\rho_i^{\text{(out)}}} = \exp\left(\beta\mu_{\text{ex},i}^{\text{(out)}} - \beta\mu_{\text{ex},i}^{\text{(in)}}\right),
\end{equation}
where $\beta \equiv 1/k_{\text{B}}T$, $k_{\text{B}}$ is the Boltzmann constant, and $T$ is the absolute temperature.
Partition coefficients are experimentally accessible and biologically relevant quantities, since they quantify the tendency of specific biomolecules to partition spontaneously into phase-separated condensates.

\subsection{Mean-field models with pairwise interactions}

The simplest theoretical descriptions of LLPS are based on mean-field models, which introduce effective parameters to describe how molecules interact with one another.
A mean-field model prescribes an approximate free-energy surface in terms of the effective interaction parameters and the component concentrations.
The most widely used mean-field models, both in the condensate literature and more generally in biophysics and materials science, make the assumption that the excess chemical potential of species $i$ can be written in the form
\begin{equation}
  \label{eq:muex}
  \mu_{\text{ex},i}(\{\rho_j\}) = \mu_{\text{v}}(\{\rho_j\}) + \beta^{-1} \sum_{j=1}^N B_{ij} \rho_j,
\end{equation}
where $\mu_{\text{v}}$ is a monotonically increasing function that depends only on the concentrations and the excluded volume associated with each molecular species.
The second term embodies the assumption of ``pairwise interactions'' among the $N$ non-solvent components, where $\{B_{ij}\}$ is an $N \times N$ symmetric matrix of interaction parameters.
This assumption underlies the regular solution model of phase-separating mixtures~\cite{porter2009phase}, the Flory--Huggins model of homopolymer phase separation~\cite{colby2003polymer}, and the van der Waals model of non-ideal fluids~\cite{hansen2013theory}.

The Flory--Huggins model~\cite{colby2003polymer} is commonly used to fit experimental data on biomolecular LLPS~\cite{qian2022analytical}.
Assuming an incompressible fluid with $N$ non-solvent species, the Flory--Huggins free-energy density is
\begin{equation}
  \label{eq:FH-free-energy}
  \beta f v_0 = \sum_{i=1}^N \frac{\phi_i}{L_i} \log \phi_i + \phi_0 \log \phi_0 + \frac{1}{2}\sum_{i=1}^N\sum_{j=1}^N \epsilon_{ij} \phi_i\phi_j,
\end{equation}
where the volume fraction occupied by species $i$ is ${\phi_i = L_i v_0 \rho_i}$, the degree of polymerization of species $i$ is $L_i$, the size of a monomer is represented by $v_0$, and the solvent-occupied volume fraction, $\phi_0$, is determined by the incompressibility constraint, $\sum_{i=0}^N \phi_i = 1$.
We note that, within the context of this model, the ``solvent'' may itself represent a mixture including non-interacting macromolecules.
The interaction parameters $\{\epsilon_{ij}\}$ are dimensionless.
Negative interaction parameters imply that molecules attract one another, while positive interaction parameters imply repulsion.
Homotypic and heterotypic interactions are encoded in the on- and off-diagonal elements of $\{\epsilon_{ij}\}$, respectively.
\eqref{eq:FH-free-energy} is consistent with \eqref{eq:muex}, since the interaction parameters only enter the free-energy density in a quadratic form.
The contribution to the free-energy density from the pairwise interactions can also be written in terms of Flory $\chi$ parameters, $\chi_{ij} = \epsilon_{ij} - (\epsilon_{ii} + \epsilon_{jj}) / 2$, by extending the sums in the final term of \eqref{eq:FH-free-energy} to include the solvent (component 0) and replacing $\epsilon_{ij}$ with $\chi_{ij}$.
This change of variables introduces terms that are linear in $\{\phi_i\}$ into the free-energy density, which have no effect on the phase behavior.
With this alternate notation, the on-diagonal elements $\{\chi_{ii}\}$ are zero by definition, and the homotypic interactions are encoded by the interactions with the solvent, $\{\chi_{i0}\}$.

Two non-solvent components are sufficient to reveal generic effects of homotypic versus heterotypic interactions.
In such a mixture, two distinct types of phase transitions can occur: A ``condensation'' transition can occur if attractive heterotypic interactions are comparable to or stronger than any attractive homotypic interactions, while a ``demixing'' transition can occur if the heterotypic interactions are significantly less attractive than one or both of the homotypic interactions~\cite{jacobs2017phase}.
Both behaviors have been observed in numerical investigations of two-component-plus-solvent mean-field (e.g.,~\cite{shek2022spontaneous}) and molecular simulation models (e.g.,~\cite{mazarakos2022multiphase}).
Condensation transitions are analogous to LLPS in simple one-component-plus-solvent fluids, implying that the phase diagram can be fully described by projecting the concentrations onto the parent composition vector~\cite{jacobs2017phase}.
By contrast, mixtures with dissimilar homotypic and heterotypic interaction strengths have more complex phase diagrams.
For example, the implications of this complexity for concentration buffering have recently been explored in \refcite{deviri2021physical} using a two-component-plus-solvent Flory--Huggins model.
Concentration buffering was shown to be effective when the tie lines connecting the coexisting condensed and dilute phases are parallel to the concentration ``noise distribution.''
This observation follows from the generalized lever rule, \eqref{eq:lever-rule}, with $K=1$, which implies that fluctuations of the parent concentrations in the direction $\vec\rho^{(1)} - \vec\rho^{(0)}$ only modify the volume fraction of the condensed phase, $x^{(1)}$, leaving the ``buffered'' concentrations of both non-solvent species in the dilute phase unchanged.

\subsection{Constructing phase diagrams of multicomponent mean-field models}
\label{sec:phase-diagrams}

Moving beyond two-solute scenarios, the construction of high-dimensional phase diagrams becomes considerably more challenging (\figref{fig:phase-diagram}).
An elegant approach for solving this problem in mixtures with up to approximately five non-solvent components was provided in \refcite{mao2019phase}.
This method exploits the fact that the common tangent plane construction is equivalent to convexification of a non-convex free-energy surface.
In this method, the free energy of a mean-field model is first evaluated at every point of an $N$-dimensional grid over the physical domain ${\phi_i \ge 0 \,\forall i}$ and ${\sum_{i=1}^N \phi_i \le 1}$.
The volume fractions at each grid point and the corresponding free-energy value constitute a single point within an $(N+1)$-dimensional space.
The convex hull of all the points within this $(N+1)$-dimensional space can then be determined using standard algorithms~\cite{barber1996quickhull}.
Importantly, grid points that lie within coexistence regions are \textit{not} part of the convex hull.
Furthermore, the facets of the convex hull can be analyzed to determine the number of coexisting phases in a coexistence region.
This algorithm can be used as ``black-box'' method for identifying coexistence regions, up to the resolution specified by the concentration-space grid, for any mean-field model.

In order to perform coexistence calculations to greater precision, it is necessary to identify the coexistence chemical potential vector that results in a grand potential with multiple global minima.
An efficient approach described in \refscite{sanders2020competing} and \cite{chen2023programmable} involves an iterative two-step algorithm.
First, assuming a fixed chemical potential vector, the local minima of the grand potential are identified using initial guesses of each of the coexisting-phase concentrations.
Then, the chemical potential vector is adjusted to bring the variance among the values of the grand potential at these local minima to zero.
This second step establishes the coexisting phases as global minima of the grand potential.
It is advantageous to use estimates of the coexisting-phase concentrations obtained from the convex-hull method as initial guesses when performing these nonlinear minimizations.
A similar approach, in which the initial guesses for the coexisting-phase concentrations are obtained from a grid-based search for the spinodal region, was proposed in \refcite{lin2022numerical}.

An alternative strategy for calculating phase coexistence has been provided in \refcite{zwicker2022evolved}.
This method uses a nonphysical dynamical scheme, inspired by swapping molecules between metastable phases, in order to eliminate differences between the chemical potentials and the pressures of the phases.
Starting from an initial guess of the component volume fractions in each of the $K+1$ coexisting phases, the dynamical scheme evolves the volume fractions in each phase $\alpha$ according to
\begin{equation}
  \label{eq:coexistence-dynamics-zwicker}
  \frac{\partial \phi^{(\alpha)}_i}{\partial t} = \phi^{(\alpha)}_i \beta \sum_{\gamma=0}^K \left[\phi_i^{(\gamma)} \!\left(\mu_i^{(\gamma)} - \mu_i^{(\alpha)}\right) + \left(P^{(\gamma)} - P^{(\alpha)}\right)\right]\!,
\end{equation}
where $\{\mu^{(\alpha)}_i\}$ and $P^{(\alpha)}$ are the component chemical potentials and the pressure, respectively, evaluated in the $\alpha$ phase with the instantaneous volume fractions $\{\phi_i\}^{(\alpha)}\!$, and $t$ is the fictitious time associated with these dynamics.
At steady state, when $\partial\phi_i^{(\alpha)}/\partial t = 0$, \eqref{eq:coexistence-dynamics-zwicker} ensures that the phases meet the thermodynamic criteria for coexistence.
Crucially, the results of this numerical approach, like the nonlinear minimization scheme described above, depend sensitively on the initial guesses for the coexisting-phase concentrations.
In particular, if a candidate phase is not represented in the $K+1$ initial concentration vectors, then it is unlikely to be captured in the final set of coexisting phases.

\subsection{Multicomponent phase coexistence in molecular simulation models via free-energy calculations}
\label{sec:free-energy-simulation}

Efficient approaches for calculating coexistence among an arbitrary number of fluid phases have also been devised for molecular simulation models.
Such models specify a potential energy function that depends on the coordinates of all particles in the simulation volume.
As such, Monte Carlo or molecular dynamics (MD) simulation methods are required to sample the configurational phase space.
A wide variety of methods are available for computing coexistence between pairs of phases~\cite{frenkel2001understanding}.
Within the condensate literature, direct coexistence simulations utilizing a ``slab geometry''~\cite{dignon2020biomolecular} have become popular due to the ease with which this approach can be implemented.
However, in order to compute phase coexistence among a larger number of phases, it is advantageous to work in the grand-canonical ensemble.
Grand-canonical phase-coexistence calculations are also ideal for minimizing finite size effects~\cite{wilding1995critical}.

\begin{figure}
  \includegraphics[width=8.5cm]{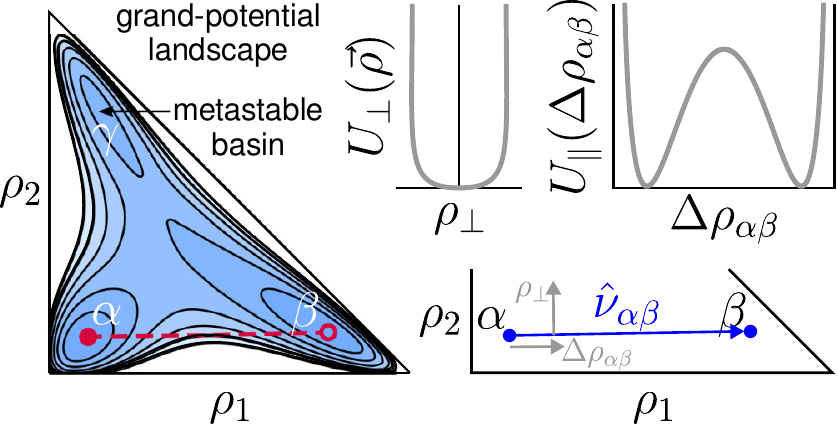}
  \caption{Multiphase coexistence points can be determined from molecular simulations by sampling the grand-potential landscape.  In order to sample two specific phases $\alpha$ and $\beta$, biasing potentials parallel, $U_\parallel$, and perpendicular, $U_\perp$, to $\hat\nu_{\alpha\beta}$ are introduced.  Reweighting techniques can then be used to tune the component chemical potentials in order to establish equal grand potentials among all coexisting phases (see text).}
  \label{fig:free-energy}
\end{figure}

A robust approach for carrying out multiphase coexistence calculations utilizes a generalization of the multicanonical sampling method~\cite{berg1992multicanonical}.
Influenced by earlier simulations of polydisperse fluids~\cite{buzzacchi2006simulation}, \refcite{jacobs2013predicting} introduced a method to sample an isolated pair of phases in a grand-canonical simulation with multiple free-energy basins (\figref{fig:free-energy}).
First, an order parameter $\Delta\rho_{\alpha\beta} \equiv (\vec\rho - \vec\rho^{(\alpha)}) \cdot \hat\nu_{\alpha\beta}$, where ${\hat\nu_{\alpha\beta} \equiv (\vec\rho^{(\beta)} - \vec\rho^{(\alpha)}) / |\vec\rho^{(\beta)} - \vec\rho^{(\alpha)}|}$, is defined to measure the distance along a linear path between the $\alpha$ and $\beta$ phases, with concentration vectors $\vec\rho^{(\alpha)}$ and $\vec\rho^{(\beta)}$, respectively.
A biasing potential is then added to constrain fluctuations in orthogonal directions of concentration space,
\begin{equation}
  U_\perp(\vec\rho) \equiv k_\perp \big|(\vec\rho - \vec\rho^{(\alpha)}) - [(\vec\rho - \vec\rho^{(\alpha)}) \cdot \hat\nu_{\alpha\beta}] \hat\nu_{\alpha\beta}\big|^{p_\perp},
\end{equation}
where $k_\perp > 0$ and $p_\perp > 0$ are user-defined constants.
An additional biasing potential in the direction of concentration space parallel to $\hat\nu_{\alpha\beta}$, $U_\parallel(\Delta\rho_{\alpha\beta})$, can then be calculated using grand-canonical Wang--Landau simulations~\cite{wang2001efficient},
\begin{equation}
  \beta U_\parallel(\Delta\rho') = \log \!\int\! d\bm{x}\, \bm{1}_{\Delta\rho_{\alpha\beta}[\vec\rho(\bm{x})], \Delta\rho'} e^{-\beta\mathcal{H}(\bm{x}) - \beta U_\perp[\vec\rho(\bm{x})]},
\end{equation}
where $\bm{x}$ represents a particle configuration, $\mathcal{H}$ is the Hamiltonian of the unbiased model, and $\bm{1}$ is the indicator function.
The biasing potential $U_\parallel$ is optimal for ``flattening'' the free-energy barrier between the $\alpha$ and $\beta$-phase regions of phase space~\cite{wang2001efficient}. 
Finally, performing a multicanonical simulation under the combined potential $\mathcal{H} + U_\perp +U_\parallel$ allows the simulation to transit reversibly between the $\alpha$ and $\beta$ phases.

\refscite{jacobs2021self} and \cite{chen2023programmable} have demonstrated how this method can be applied to calculate multiphase coexistence points for multicomponent lattice models.
Samples obtained from multicanonical simulations between different pairs of phases can be combined via reweighting methods such as MBAR~\cite{shirts2008statistically} as long as one of the phases is sampled in every simulation.
Grand potential differences between all pairs of phases can then be determined, and the chemical potentials can be adjusted in order to find the coexistence point at which all phases have identical pressures at equilibrium.
This approach has been successfully applied to compute coexistence points involving more than five phases.
Nonetheless, this method also requires prior knowledge of the approximate concentrations of all phases in order to construct the required biasing potentials and sample all the coexisting phases.

\section{Predicting and designing phase behavior in multicomponent fluids}
\label{sec:mean-field}

We now turn to theoretical studies of mixtures governed by pairwise interactions.
We first discuss efforts to predict phase behavior in mixtures with hundreds or thousands of components based on the statistical properties of the pairwise interactions.
We then describe recently devised methods to design or ``evolve'' pairwise interactions in order to stabilize a target phase diagram.

\subsection{Multicomponent mixtures with random pairwise interactions}
\label{sec:random-mixtures}

Pairwise interaction models, due to their simplicity, are a natural place to begin exploring how the presence of many distinct molecular components influence the phase behavior of a mixture.
However, theoretical progress cannot be made without specifying the form of the interaction matrix, and limited systematic experimental data exist for parameterizing heterotypic interactions.
To deal with this lack of information, \refcite{sear2003instabilities} proposed that the pairwise interactions can be modeled using a random matrix.
Specifically, \refcite{sear2003instabilities} considered symmetric random matrices in which the elements are chosen independently from a Gaussian distribution with a prescribed mean and standard deviation.
An ensemble of ``random mixtures'' is thus associated with a particular Gaussian distribution and the number of distinct components $N$, such that each mixture in the ensemble is defined by a particular realization of the $N \times N$ interaction matrix.

\begin{figure}
  \includegraphics[width=8.5cm]{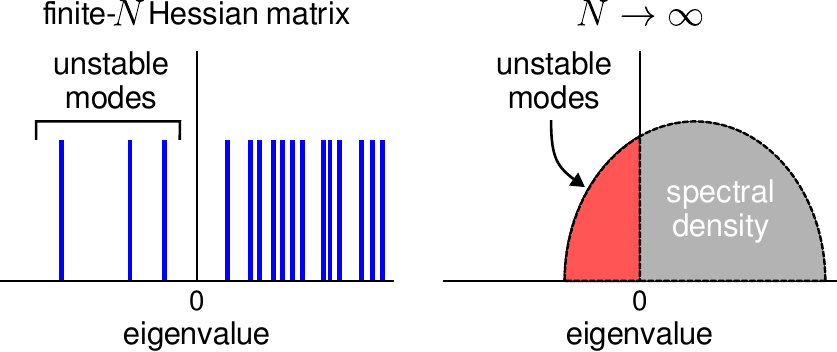}
  \caption{The spinodal locus, where the mixture becomes unstable with respect to concentration fluctuations, can be predicted using a linear stability analysis.  \textit{Left:} Computing the eigenspectrum of the Hessian matrix, $\partial^2 f/\partial\rho_i\partial\rho_j$, at the parent concentrations reveals the number of unstable modes, each of which is associated with an orthogonal direction in concentration space.  \textit{Right:} Analytical predictions in the large-$N$ limit provide insight into the relationship between the structure and statistical properties of an interaction matrix and the phase behavior of the associated biomolecular mixture.}
  \label{fig:spectral}
\end{figure}

\refcite{sear2003instabilities} assumed for simplicity that the mixture free-energy density can be described by \eqref{eq:muex} with $\mu_{\text{v}} = 0$.
The resulting free-energy density, $f$, is applicable to solutions in which all components are present at low concentrations, and the $\{B_{ij}\}$ elements in \eqref{eq:muex} are referred to as second-virial coefficients~\cite{hansen2013theory}.
By restricting the study to mixtures with equimolar parent concentrations, $\rho_i^{(\text{parent})} = \bar\rho^{(\text{parent})}\,\forall i$, it was shown that the spinodal locus can be predicted directly from the second-virial matrix.
The central idea is that unstable concentration fluctuations can be determined from a linear stability analysis of the mean-field free-energy landscape (\figref{fig:spectral}).
With the equimolar parent-concentration assumption, the eigenvalue spectrum of the Hessian matrix, $\partial^2 f/\partial\rho_i\partial\rho_j$, is equal to the spectrum of $\{B_{ij}\}$ plus a constant $1/\bar\rho^{(\text{parent})}\!$.
Instabilities therefore occur when the minimum eigenvalue of $\{B_{ij}\}$ is less than $-1/\bar\rho^{(\text{parent})}\!$.
Applying results from random matrix theory, it was shown that the existence and nature of the dominant instability, which coincides with the minimum eigenvalue of the Hessian matrix, can be determined from the mean, $b$, and standard deviation, $\sigma$, of the Gaussian distribution of matrix elements in the limit of large $N$.
Two distinct cases were observed.
If the standard deviation among the matrix elements is sufficiently small, such that $N^{1/2}b/\sigma \lesssim -1$, then the dominant instability involves concentration fluctuations that are parallel to the equimolar parent concentration vector.
This type of instability is consistent with a condensation transition driven by similar homotypic and heterotypic interaction strengths.
By contrast, if the standard deviation among the matrix elements is sufficiently large, such that $N^{1/2}b/\sigma \gtrsim -1$, then the dominant instability is orthogonal to the parent concentration vector, and individual components demix into phases with differing compositions.
Importantly, these behaviors are self-averaging, meaning that the tendency of any particular random-mixture realization to undergo a condensation or demixing transition converges in probability as $N \rightarrow \infty$.

\refcite{thewes2022composition} extended these results to mixtures with non-equimolar parent compositions.
This work considered a regular-solution free-energy density, in which ${\mu_{\text{v}} = -\log \rho_0}$ in \eqref{eq:muex}.
This additional contribution to the free energy accounts for the entropy of the solvent, providing a better physical model of solutions at non-dilute concentrations.
Modifying the free-energy density in this way does not qualitatively alter the conclusions of \refcite{sear2003instabilities} regarding condensation and demixing in equimolar mixtures.
However, consideration of non-equimolar parent compositions reveals a third type of spinodal instability: Demixing transitions can now be classified as either ``random,'' in which all the components of the eigenvector associated with the instability are of similar order, or ``localized,'' in which the demixing transition is dominated by only a few species.
Random mixtures with a large interaction-parameter variance and equimolar parent compositions tend to undergo random demixing.
By contrast, mixtures in which one component has a much higher parent concentration than all the others can undergo a composition-driven transition, in which demixing is localized to the dominant species.
The authors emphasized that the direction of composition-driven instabilities cannot be predicted simply by considering the relative parent concentrations of the components; instead, the interplay between entropic effects and random pairwise interactions tends to amplify the contribution of the dominant component to the unstable concentration fluctuations.
In other words, the nature of the instabilities at the spinodal locus of a random mixture depends on both the interaction matrix and the parent concentrations.

Simulation support for the qualitative predictions of \refcite{sear2003instabilities} was provided in \refscite{jacobs2013predicting} and \cite{jacobs2017phase}.
In these studies, the free-energy calculation strategy described in \secref{sec:free-energy-simulation} was applied to compute coexistence between an equimolar dilute phase and a condensed phase in random mixtures with up to 64 non-solvent components.
Simulations were conducted using a multicomponent lattice model, with the nearest-neighbor interactions between particles on the lattice specified by a random interaction matrix generated according to the Gaussian prescription of \refcite{sear2003instabilities}.
Coexistence calculations were then performed to investigate the nature of the phase transition that occurs at the lowest total parent concentration, meaning that the simulated coexistence point represents the lowest-concentration intersection of the equimolar parent concentration vector with any coexistence region.
The average phase behavior of the random-mixture ensemble was analyzed by repeating these calculations for many independent realizations of random mixtures with the same interaction mean and variance.

Although the lattice-based coexistence calculations of \refscite{jacobs2013predicting} and \cite{jacobs2017phase} are not directly comparable to theoretical predictions regarding instabilities at the spinodal locus, analogous condensation and demixing transitions were observed in this molecular simulation model.
First, the phase behavior at each simulated coexistence point was classified as condensation or demixing according to the angle, $\theta$, between the equimolar parent concentration vector and the unit vector connecting the coexisting phases, $\hat\nu_{\alpha\beta}$.
This angle was found to be self-averaging with respect to the number of components, $N$, as suggested by random matrix theory~\cite{jacobs2013predicting}.
Second, \refcite{jacobs2017phase} observed that the distribution of $\theta$ is bimodal, signifying a sharp transition between these two qualitatively distinct types of phase transitions as the mean and/or variance of the random-interaction distribution was changed.
Third, increasing the number of components was found to shift the phase behavior at the simulated coexistence points towards condensation transitions, in line with the predictions of \refcite{sear2003instabilities}.
This finding implies that the mixing entropy of multicomponent fluids acts to suppress demixing instabilities.
However, by contrast with \refcite{sear2003instabilities}, simulation results indicated that the extreme values of the interaction matrix are more predictive of the simulated coexistence concentrations than the eigenspectrum of the mean-field Hessian matrix.
This observation was exploited to propose a scaling relation for the transition between condensation and demixing behaviors at the phase boundary, $(\log N)^{1/2} \sim \sigma$, that differs from the random-matrix-theory prediction for the condensate--demixing crossover at the spinodal locus, $N^{1/2} \sim \sigma/b$.
This idea has since been followed up in \refcite{girard2022kinetics}, which suggested that the coexistence points can be strongly influenced by the tails of the distribution from which the elements of the random interaction matrix are chosen.

Phase separation in mean-field models of mixtures with many components has also been analyzed using phase-field simulations~\cite{shrinivas2021phase}.
Deterministic phase-field simulations evolve the spatially varying component volume fractions, $\{\phi_i(\vec r)\}$, on a three-dimensional grid in accordance with linear irreversible thermodynamics~\cite{onsager1931reciprocal}.
As such, phase-field simulations reach a steady state when the free energy of the simulated volume reaches a local minimum; this steady state may be spatially inhomogeneous if phase separation occurs.
\refcite{shrinivas2021phase} considered a regular-solution free-energy density consistent with \eqref{eq:muex}, with ${\mu_{\text{v}} = -\log\phi_0(\vec r) - \kappa \nabla^2\phi_i (\vec r)}$.
The second term in $\mu_{\text{v}}$, which penalizes the formation of interfaces between phases in a component-independent manner, arises from square-gradient contributions to a Cahn--Hilliard free-energy functional with $\kappa > 0$~\cite{cahn1958free}.
Simulations then implemented ``Model B dynamics''~\cite{provatas2011phase}, where ${\partial\phi_i/\partial t = \nabla \cdot (M \phi_i \nabla \mu_i)}$, with a component-independent mobility coefficient $M > 0$.
Upon reaching steady state, compositionally distinct phases were identified by performing a principal component analysis of the spatially varying component concentrations.

Since phase separation in a deterministic phase-field model proceeds via spinodal decomposition, \refcite{shrinivas2021phase} was able to provide a direct test of the analytical predictions of \refcite{sear2003instabilities}.
Both condensation and demixing were observed in simulations initialized with equimolar parent concentrations.
Consistent with a linear stability analysis at these initial conditions (\figref{fig:spectral}), \refcite{shrinivas2021phase} found that the number of phases identified at steady state correlates with the number of negative eigenvalues of the Hessian matrix.
Furthermore, the number of steady-state phases could be estimated from the limiting ($N\rightarrow\infty$) spectral density predicted by random matrix theory.
This trend was shown to hold for a variety of random-mixture ensembles in which the standard deviation of the independently sampled interaction-matrix elements was either held constant or scaled proportionally to $N^{1/2}$.
Nonetheless, some caution is warranted in interpreting these results, since the steady-state found via spinodal decomposition may reflect a metastable configuration that does not represent all the equilibrium phases.
We shall return to this important consideration below in \secref{sec:iterative-design}.

\subsection{Multicomponent mixtures with structured pairwise interactions}
\label{sec:structured}

Although random-mixture models are useful for investigating generic features of high-dimensional phase diagrams, they may not reflect the structure of pairwise interactions among real biomolecules.
In particular, the assumption that the elements of a $\{B_{ij}\}$ matrix are independently and identically distributed implies that $\mathcal{O}(N^2)$ pairwise coefficients characterize the mixture, even though there are only $N$ chemically distinct biomolecules.
Physical interactions arising from the physicochemical features of the biomolecules are instead likely to introduce correlations into the $\{B_{ij}\}$ matrix.

To address this critical issue, ``structured'' pairwise interaction models have been introduced and studied using linear stability analysis.
\refcite{carugno2022instabilities} took the approach of grouping components into distinct families, whereby all members within a particular family have similar physicochemical properties.
The authors proposed that this relationship could be described via by an interaction matrix of the form ${B = D + C \ast Z}$, where $\ast$ indicates element-wise multiplication.
$D$ and $C$ are block matrices specifying the mean and standard deviation of the interactions between families, respectively, while $Z$ is a Gaussian random matrix with zero mean and unit variance.
This model reduces to the random-mixture model of \refcite{sear2003instabilities} when there is only one family, in which case all interactions have the same mean and variance.
Intuitively, a single family of components can demix from a mixture with equimolar parent concentrations if the intra-family interactions are sufficiently more attractive than inter-family interactions.
Such ``family demixing'' tends to dominate over random demixing when the noise amplitude, governed by $C$, is small.

\refcite{graf2022thermodynamic} explored an alternative approach in which structured interaction matrices are assumed to have a low matrix rank.
This assumption implies that the interaction matrix can be written in the form ${B_{ij} = \sum_{l=1}^r c^{(l)} s_i^{(l)} s_j^{(l)}}$, where the index $l$ is bounded by the matrix rank, $r$.
This low-rank decomposition was inspired by a toy model in which each molecular species can be described by $r$ ``molecular features,'' which interact according to diagonalized coupling coefficients $\{c^{(l)}\}$.
The matrix $\{s_i^{(l)}\}$ specifies the value of each molecular feature for each component $i$.
In fact, any $N \times N$ interaction matrix can be written in this form via eigendecomposition, assuming that $N-r$ of its eigenvalues are negligible.
If all the nonzero eigenvalues of $\{B_{ij}\}$ are negative, representing net attractive interactions among the molecular features, then the linear-stability condition for the spinodal locus can be recast in terms of a feature covariance matrix.
Specifically, this rank-$r$ matrix measures the covariance among the values of the molecular features, weighted by the concentrations of the components expressing these features, in a homogeneous mixture with fixed parent concentrations.
The directions of the unstable concentration fluctuations can then be determined from the first principal component of the concentration-weighted molecular-feature distribution.
This result bears resemblance to related studies of polydisperse fluids, in which phase transitions have been predicted using so-called ``moment free energies''~\cite{sollich2001predicting,sollich2001moment}.
When $\{B_{ij}\}$ has both positive and negative eigenvalues, covariance matrices for the net-attractive and net-repulsive molecular features must be considered separately.
The extent to which the net-repulsive features modify the phase behavior depends on whether their concentration-weighted distribution correlates with that of the net-attractive feature distribution.
The authors also showed that this analysis can be extended to predict ordinary and higher-order critical points, whose occurrence depends on higher-order cumulants of the concentration-weighted feature distribution.

An important insight gained from this theory~\cite{graf2022thermodynamic} is that the phase behavior of a mixture can be predicted by analyzing properties of the $r$-dimensional feature space, which may be much simpler than the $N$-dimensional concentration space if $r \ll N$.
Since intermolecular interactions among conformationally disordered biomolecules are widely believed to arise from a limited number of chemical interactions, such as electrostatic interactions among charged amino acids and hydrophobic forces involving amino acids with aromatic side chains, it is plausible that this is indeed the case.
The relationship between this ansatz and findings from sequence-dependent theories will be discussed in \secref{sec:sequence-dependent}.
The work of \refcite{graf2022thermodynamic} has also suggested a useful method for coarse-graining a multicomponent fluid into an equivalent binary mixture with the same spinodal and critical points by preserving the second and third cumulants along the first principal component of the concentration-weighted feature distribution.
However, it is unclear whether the coexistence manifolds of multicomponent mixtures with low-rank interaction matrices can be simplified in the same way.

\subsection{Iterative design of multicomponent phase behavior}
\label{sec:iterative-design}

Taking the next step towards biologically realistic mixtures requires consideration of specific interactions that have emerged due to evolutionary processes.
Recent efforts~\cite{jacobs2021self,zwicker2022evolved,chen2023programmable} to explore the thermodynamic consequences of evolved interaction specificity have shown that multicomponent mixtures can be designed with the goal of stabilizing a prescribed number of condensed phases.
The logic behind this approach is that immense size of the space of possible biomolecular interactions limits the probability that a random-mixture model will produce a phase diagram comparable to the observed complexity of intracellular phase-separated condensates.
Indeed, even in the simplest pairwise-interaction models, the ``design space'' has a dimension of $N(N+1)/2$ when all interactions are independently controllable.
By contrast, treating multicomponent LLPS as an optimization problem in which the interactions can be systematically tuned has the potential to discover regions of this design space that are relevant to multiphasic condensates.

\refcite{zwicker2022evolved} demonstrated that the number of coexisting phases in a mean-field pairwise-interaction model can be designed by iterative application of a genetic algorithm.
This design process necessitates finding all coexisting phases given a candidate interaction matrix at each iteration.
The genetic algorithm is then applied to evolve a population of interaction matrices in order to identify matrices that result in a target ``phase count'' of condensed phases.
It turns out that this goal is surprisingly easy to achieve owing to the size of the design space when all pairwise interactions are independently tunable.
An intuitive strategy of designing block-diagonal matrices, along the lines of \refcite{carugno2022instabilities}, reliably results in phase counts equal to the number of blocks of strongly attractive interactions.
However, the genetic algorithm finds solutions to this design problem that are less obviously structured.
The authors further showed that designed mixtures with low phase counts tend to be stable with respect to small random perturbations in the interaction energies and that the genetic algorithm can rapidly alter the phase count of a designed mixture, finding new solutions within a few tens or hundreds of iterations.

This iterative design approach comes with a number of caveats, however.
First, optimizing for a target phase count does not guarantee that different solutions identified by the genetic algorithm correspond to condensates with similar molecular compositions.
Second, although the phase count of a candidate interaction matrix should depend on the parent concentrations according to \eqref{eq:lever-rule}, \refcite{zwicker2022evolved} employed a strategy of sampling coexistence points at random parent concentrations.
This approach suggests an implicit design goal of maximizing the volume of the $(K+1)$-phase coexistence region within the $N$-dimensional concentration space.
Third, the reliability and performance of the iterative design algorithm are sensitive to the computational cost and accuracy, respectively, of the intermediate phase-coexistence calculations, which must be repeated for each candidate interaction matrix.
This is in fact a very general problem: Regardless of the mixture model, phase-coexistence calculations first require a search for candidate phases, whether by exhaustive grid-based sampling (e.g.,~\cite{mao2019phase}; see \secref{sec:phase-diagrams}), randomized initial conditions (e.g.,~\cite{zwicker2022evolved}; see \secref{sec:phase-diagrams}), Monte Carlo sampling (e.g.,~\cite{jacobs2017phase}; see \secref{sec:free-energy-simulation}), or physical dynamics (e.g.,~\cite{shrinivas2021phase}; see \secref{sec:random-mixtures}).
The computational cost of this search problem scales exponentially with the dimension of the concentration space.

\subsection{Inverse design of multicomponent phase behavior}
\label{sec:inverse-design}

\begin{figure}
  \includegraphics[width=8.5cm]{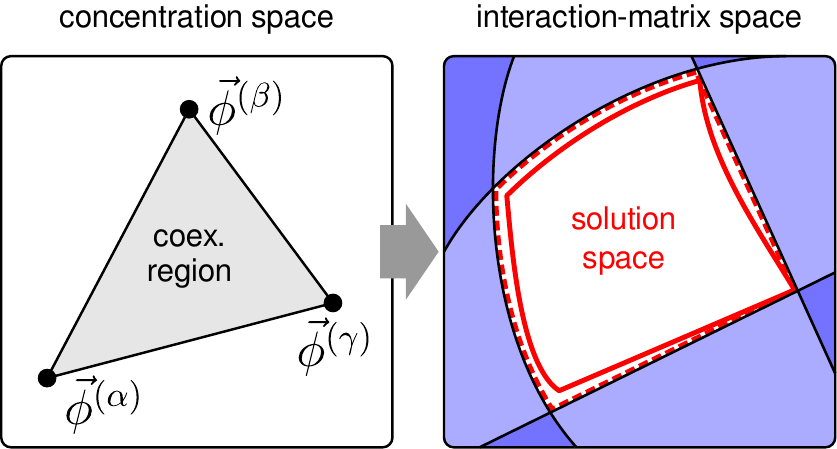}
  \caption{In the inverse design approach, restrictions on the solution space of pairwise interaction matrices are determined directly from the concentrations of the target phases and the thermodynamic criteria for phase coexistence.  \textit{Left:} The target phase diagram consisting of condensed phases $\{\vec\phi^{(\alpha)}\}$.  Any mixture with parent concentrations inside the convex hull of the target phases will phase-separate at equilibrium to establish coexisting phases with the prescribed concentrations.  \textit{Right:} Convex programming can be applied to compute the subspace containing interaction matrices that are consistent with the target phase behavior.  The convex volume (red dashed line) bounded by the convex-optimization constraints (black lines) closely approximates the solution space to the inverse problem (red solid line).  Because many interaction matrices may yield the same phase behavior, regularization is needed to select a particular matrix from the solution space.}
  \label{fig:inverse-design}
\end{figure}

Many of the drawbacks of iterative design approaches can be overcome by directly solving the \textit{inverse problem}---designing interactions to yield target phase behavior.
Inverse design entails working out constraints on the solution space of biomolecular interactions that correspond to desired collective properties, such as the compositions of condensed phases (\figref{fig:inverse-design}).
Suitable interactions can be identified in this way without explicitly performing phase-coexistence calculations.
As a result, the computational requirements may scale more favorably with the number of components, in particular because the initial search for candidate phases can be avoided.

An inverse design strategy for mixtures with pairwise interactions was first introduced in \refcite{jacobs2021self}.
Because \eqref{eq:muex} is linear with respect to $\{B_{ij}\}$, the inverse problem can be solved approximately using a convex relaxation.
It is therefore possible to prove, within the convex relaxation, whether a pairwise interaction matrix exists for a prescribed set of immiscible phases, and if so, to calculate a suitable interaction matrix with efficient convex programming algorithms~\cite{boyd2004convex}.
\refcite{jacobs2021self} showed that the thermodynamic requirements for establishing metastable phases with prescribed compositions yield a convex relaxation known as a semidefinite program (SDP).
The SDP constraints comprise both affine and eigenvalue inequalities, since the Hessian matrix must be positive definite in each target phase.
Solutions to this SDP were shown to result in metastable phases with the desired compositions in mixtures with up to 200 distinct components, both in the context of a Flory--Huggins mean-field model and in Monte Carlo simulations of an associated multicomponent lattice model.

Exploiting the ability to prove feasibility of the SDP, \refcite{jacobs2021self} then studied the probability of finding a feasible solution for an inverse problem with randomly assigned target-phase compositions.
This probability was found to drop sharply beyond a certain number of target phases, revealing a thresholding transition reminiscent of the storage capacity in the Hopfield model of neural networks~\cite{hopfield1982neural} and ``multifarious'' self-assembly of finite-sized structures~\cite{murugan2015multifarious}.
The critical number of condensed phases associated with this thresholding transition could be predicted using graph-theoretic arguments and depends on both the number of components in the mixture and the fraction of components whose concentrations are enriched in each target phase relative to the surrounding fluid.

A similar convex optimization approach was then applied to design mixtures with prescribed equilibrium phases~\cite{chen2023programmable}.
A two-step procedure for designing pairwise interaction matrices was proposed.
First, a convex relaxation was used to specify an SDP for both the interaction matrix and the approximate coexistence chemical potential vector.
Then, the chemical potentials were adjusted to ensure coexistence among the target phases using the nonlinear algorithm described in \secref{sec:phase-diagrams}.
A regularization heuristic was also introduced to pick out a unique interaction matrix from within the solution space, eliminating competing condensed phases that were not specified in the phase-diagram design problem.
Applying this approach to the Flory--Huggins model, \eqref{eq:FH-free-energy}, \refcite{chen2023programmable} provided numerical evidence that while the feasibility of the SDP is independent of the degree of polymerization, the convex relaxation becomes a better approximation of the phase-diagram design problem as the degree of polymerization increases (\figref{fig:inverse-design}).
Interestingly, coexistence regions with more condensed phases than distinct mixture components can be designed in this way.
Furthermore, this inverse design approach is easily extended to include additional optimization goals or constraints on the interactions; for example, it is possible to compute the minimum number of matrix-elements that must be changed in order to switch from one phase diagram to another using this method.
\refcite{chen2023programmable} also demonstrated that by mapping interaction matrices to molecular pair potentials, interactions designed using mean-field models can be used to establish coexistence among phases with prescribed compositions in molecular simulation models.

In another application of inverse design, \refcite{mao2020designing} devised an algorithm to engineer pairwise interactions that produce phase-separated condensates with target morphologies, such as those observed in the nucleolus~\cite{feric2016coexisting}.
At equilibrium, surface tensions control the tendency of macroscopic droplets to exist in nonwetting, partial wetting, or complete wetting configurations (separated, fused, and enveloped droplets, respectively, in \figref{fig:biomolecules}).
Furthermore, within the Cahn--Hilliard framework~\cite{cahn1958free}, the surface tensions between phases of mutually immiscible components are directly related to the pairwise interactions.
\refcite{mao2020designing} showed that predicting multiphase morphologies in multicomponent fluids corresponds to a graph decomposition problem, in which vertices indicate phases and edges indicate shared interfaces between phases.
Designing interaction matrices for multicomponent mixtures that phase separate into droplets with prescribed (non)wetting architectures can therefore be achieved by encoding the desired morphology in a graph, enumerating affine inequality constraints on the interactions via graph decomposition, and solving the resulting linear program.
Phase-field simulations were then used to demonstrate the efficacy of this design algorithm.

\section{Sequence-dependent theories and coarse-grained molecular models}
\label{sec:sequence-dependent}

\begin{figure}
  \includegraphics[width=8.5cm]{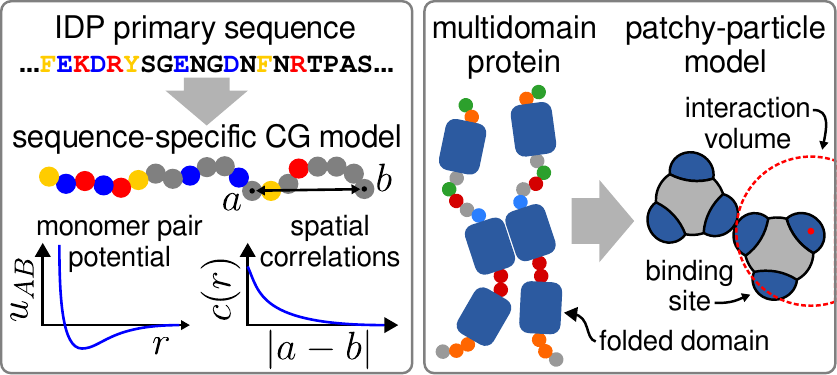}
  \caption{\textit{Left:} Sequence-dependent CG models represent IDPs as chains of simplified amino acids.  Typically, the nonbonded interactions between amino acids of types $A$ and $B$ are modeled using a pair potential, $u_{\text{AB}}(r)$.  Analytical theories of sequence-dependent heteropolymer interactions also require a model of the spatial correlations between monomers that are spaced a distance $|a-b|$ apart in the primary sequence.  \textit{Right:} Patchy-particle models of multidomain proteins implement a higher level of coarse-graining by treating the PPI or RNA-binding interfaces on folded domains as specific binding sites on simplified particles; binding sites engage in at most one interaction at a time.  Analytical theories associate an interaction volume with each pair of distinct binding-site types.}
  \label{fig:sequence-dependent}
\end{figure}

In parallel with efforts to understand the phase behavior of simplified mixtures with many components, theoretical models have been developed to describe LLPS at a greater level of chemical detail in solutions with a small number of distinct biomolecular species (\figref{fig:roadmap}).
In the condensate literature, such models can be broadly classified as sequence-specific coarse-grained (CG) IDP models, which represent nonbonded interactions between amino acids~\cite{dignon2018sequence,dignon2018relation,latham2019maximum,joseph2021physics,tesei2021accurate,tesei2022improved} or chemical functional groups~\cite{benayad2020simulation} using pair potentials, and ``patchy-particle''~\cite{jacobs2014phase,espinosa2019breakdown,joseph2021thermodynamics} or ``patchy-polymer''~\cite{zhang2021decoding,murata2022stoichiometric} CG models, which encode specific interactions between discrete binding sites on each molecule (\figref{fig:sequence-dependent}).
We first describe key insights into multicomponent phase behavior from theoretical analyses of these types of models before reviewing recent multicomponent molecular simulation studies.

\subsection{Multicomponent field-theoretic approaches}
\label{sec:field-theoretic}

Field-theoretic approaches have been used to predict the sequence-dependent phase diagrams of heteropolymers, with a particular emphasis on polyampholytes.
By accounting for chain connectivity, and thus the primary sequence of the heteropolymer, these approaches improve upon mean-field treatments that consider all monomer--monomer interactions in a polymer solution independently~\cite{lin2018theories}.
Field-theoretic approaches incorporate sequence information by modeling the correlations between monomers within a single chain, which decay with increasing separation between monomers along the primary sequence (\figref{fig:sequence-dependent}).

\refscite{lin2016sequence} treated the spatial correlations between monomers with the random-phase approximation (RPA) by assuming that the polymer configurations obey the Gaussian statistics of ideal chains.
This assumption means that monomers on different chains are not spatially correlated and that the heteropolymer sequences affect the potential energy, but not the polymer conformations, of a mixture at finite concentration.
Despite this simplification, RPA predictions correlate well with experimental measurements of the phase behavior~\cite{lin2016sequence} and single-chain properties~\cite{lin2017phase} of charge-neutral polyampholytes.
Of particular importance, the RPA theory rationalizes the observed increase in LLPS propensity of charge-neutral sequences with ``blocky'' as opposed to homogeneous charge patterns~\cite{mccarty2019complete}.
Blocky charge patterns also correlate with smaller radii of gyration of chains in the dilute phase, in line with prior studies using the ``sequence charge decoration'' order parameter~\cite{sawle2015theoretical,firman2018sequence} and related blockiness metrics~\cite{mao2010net,das2013conformations}.
The RPA theory was extended to charged polyelectrolytes in \refcite{lin2020unified}.

Of relevance to multicomponent mixtures, \refcite{lin2017charge} applied RPA to mixtures of two distinct charge-neutral polyampholytes.
Because RPA ignores spatial correlations between monomers on different chains, the electrostatic contribution to the RPA free energy can be factored into terms arising from each chain individually.
The RPA free energy can therefore be mapped at low concentrations to a pairwise-interaction model in which the effective heterotypic interaction, $B_{12}$, is the geometric mean of the two homotypic interactions, $B_{11}$ and $B_{22}$.
The homotypic interaction coefficients can be calculated by applying RPA to each heteropolymer sequence individually.
In light of the discussion in \secref{sec:structured}, these results indicate that RPA predicts a rank-1 pairwise interaction matrix for charge-neutral polyampholyte mixtures, since ${B_{ij} = B_{ii}^{1/2}B_{jj}^{1/2}\,\forall i,j}$.
This observation further suggests that spatial correlations between different chains are needed to predict higher-rank interaction matrices for charge-neutral polyampholytes.

RPA has also been applied to mixtures of polyelectrolyte mixtures.
In solutions with two positively charged and one negatively charged polymer, \refcite{chen2021multiphase} predicted that multiphase coacervates can form due to the repulsive heterotypic interactions between two positively charged sequences with differing linear charge densities.
\refcite{chen2023multiphase} then predicted that differences in the charge patterning between two positively charged sequences with identical linear charge densities is sufficient to drive the formation of two immiscible condensed phases.

An analogous field-theoretic treatment of heteropolymers interacting via short-ranged hydrophobic forces revealed that the leading order contribution to the interaction free energy is given by the sum of the interactions between all pairs of monomers in the mixture~\cite{wessen2022analytical}.
Although this model was not explicitly applied to multicomponent solutions, it suggests that, to leading order, the pairwise interaction matrix for heteropolymers interacting via short-ranged interactions is independent of their primary sequences.
In other words, only the frequency of each monomer type in a heteropolymer sequence is relevant at this level of theory~\cite{de2022phase}, and the rank of the interaction matrix cannot exceed the rank of the monomer--monomer interaction matrix, which may itself be rank-deficient~\cite{wessen2022analytical,li1997nature}.

\subsection{Multicomponent associating fluid models}
\label{sec:associating-fluid}

Concepts from associating fluid theory~\cite{wertheim1984fluids,chapman1989saft} have been adopted to describe the interactions between binding sites on biomolecules that can only engage in one physical bond at a time.
While the methods of \refscite{wertheim1984fluids} and \cite{chapman1989saft} were originally developed to describe site-specific associative interactions between small molecules, this physical picture extends naturally to multidomain proteins or protein complexes whose constituent domains contain interfaces that interact specifically with other proteins or RNA sequences~\cite{jacobs2014phase,jacobs2016oligomers}.
The number of such binding sites therefore establishes the coarse-grained ``valence'' of the multidomain protein or complex (\figref{fig:sequence-dependent}).

Associating fluid theory treats the attractive interactions between pairs of binding sites as perturbations to the free energy of a reference model, which represents the molecular mixture in the absence of binding sites.
For example, the Flory--Huggins homopolymer model can serve as a reference model for a mixture of multidomain proteins, with the degree of polymerization $L_i$ taken to be equal to the number of domains in each protein species~$i$~\cite{jacobs2014phase}.
The concentration-dependent site--site binding probabilities are then determined from the chemical equilibrium equations
\begin{equation}
  \label{eq:saft-chem-eq}
  X_{iA} + X_{iA} \sum_{j=1}^N \rho_j \sum_{B=1}^{m_j} X_{jB} \Delta_{iA,jB} = 1\quad\forall i,A,
\end{equation}
where $X_{iA}$ represents the probability that the binding site of type $A$ on a molecule of type $i$ is not engaged in any associative interaction, and $m_i$ is the valence of molecule type $i$.
The matrix $\{\Delta_{iA,jB}\}$ represents the interaction volumes (i.e., the reciprocals of the dissociation constants) for the associative interactions between binding sites $A$ and $B$, which can in principle depend on spatial correlations in the reference model.
Finally, the contribution to the free-energy density due to associative interactions is~\cite{chapman1989saft}
\begin{equation}
  \label{eq:saft-fe}
  \beta f_{\text{assoc}} = \sum_{i=1}^N \left[ \rho_i \sum_{A=1}^{m_i} \left(\log X_{iA} - \frac{X_{iA}}{2}\right) + \frac{m_i}{2} \right].
\end{equation}
\refcite{michelsen2001physical} showed that \eqref{eq:saft-chem-eq} has a unique solution and that \eqref{eq:saft-fe} leads to a particularly simple expression for the associative contribution to the excess chemical potential when $\Delta_{iA,jB}$ is concentration-independent, ${\beta\mu_{\text{assoc},i} = \sum_{A=1}^{m_i} \log X_{iA}}$.
Furthermore, in the limit of weak associative interactions, \eqref{eq:saft-fe} reduces to a simple pairwise form, such that ${\beta\mu_{\text{assoc},i} \simeq -\sum_{j=1}^N \rho_j \sum_{A=1}^{m_i} \sum_{B=1}^{m_j} \Delta_{iA,jB}}$.
With regard to the discussion in \secref{sec:structured}, the maximum rank of the pairwise interaction matrix is therefore given by the rank of $\{\Delta_{iA,jB}\}$ in this limit.

The associating fluid framework has been widely applied to model biomolecular LLPS involving folded domains that interact via specific binding sites.
A notable application of the associating fluid framework to multiphase condensates was provided in \refcite{sanders2020competing}, which used a simplified representation of an experimentally determined PPI network to predict the compositions and morphologies of coexisting stress granule and P-body condensates.
Agreement between theory and experiment regarding the effects of concentration changes and binding-site modifications provided strong evidence that the phase behavior of these condensates is indeed governed by specific PPIs and interactions between RBDs and mRNA.

When the binding sites are assumed to represent individual amino acids of IDPs or short sequence motifs of IDPs and/or RNAs, associating fluid theory is commonly referred to as the ``stickers-and-spacers'' model of heteropolymer association~\cite{semenov1998thermoreversible,tanaka2002theoretical,choi2020physical,michels2021role}.
In this case, a Flory--Huggins homopolymer model with a degree of polymerization much greater than the binding-site valence (i.e., the number of ``stickers'') is typically taken as the reference model.
Stickers-and-spacers applications of associating fluid theory have been successfully used to rationalize experimental observations of IDP-driven phase separation, including both thermodynamic and dynamical properties, in many contexts~\cite{choi2020physical,wang2018molecular,martin2020valence}.
The assignment of the ``stickers'' to specific amino acids or short sequence motifs has varied depending on context across different studies, however, suggesting that additional contextual information may be required to predict the phase behavior of multicomponent IDP and RNA mixtures from their sequences.
For further discussion of applications of associating fluid theory to biopolymers, we direct the reader to recent reviews on this subject, including \refscite{choi2020physical} and \cite{pappu2023phase}.

\subsection{Insights from coarse-grained molecular simulations of multiphase condensates}

\subsubsection{Polymer simulations with pair potentials}

Molecular simulations have provided insights into the accuracy of analytical theories for describing sequence-dependent multicomponent phase behavior.
In order to test the RPA predictions of \refcite{lin2017charge} (see \secref{sec:field-theoretic}), \refcite{pal2021subcompartmentalization} used a combination of field-theoretic and CG MD simulations to study the phase behavior of polyampholyte mixtures.
These simulations demonstrated that pairs of charge-neutral sequences only exhibit demixing when the chains have sufficiently different (i.e., blocky versus uniform) charge distributions.
These results are in line with the predictions of the RPA theory.
Nonetheless, the authors found that excluded volume interactions---which are present in the MD simulations but are not included in the RPA calculations---are essential for observing demixing in MD simulations.
The qualitative agreement with the theoretical predictions was therefore ascribed to the assumption of incompressibility in the RPA calculations.
Nonetheless, this observation points to the need for more accurate theoretical treatments that account for excluded volume and interchain correlations.

Moving to systems with a third non-solvent component, \refcite{kaur2021sequence} performed simulations of a three-component system comprising a prion-like polypeptide (PLP), an arginine-rich polypeptide (RRP), and RNA.
In this system, competition between PLP and RNA for binding to RRP results in the demixing of PLP+RRP condensates into immiscible PLP and RNA+RRP phases when RNA is added.
This experimental observation, which bears qualitative resemblance to the competing heterotypic model of \refcite{chen2021multiphase} (see \secref{sec:field-theoretic}), was reproduced using MD simulations of a CG IDP/RNA model.
These simulations also rationalized the experimental observation that the RNA parent concentration controls the morphology of the coexisting condensates.

In an attempt to uncover general sequence determinants of multiphase mixtures, \refcite{chew2023thermodynamic} proposed a computational approach to design IDP sequences that result in multilayered condensates.
To this end, the authors used a genetic algorithm to optimize pairs of sequences that form immiscible phases and a stable shared interface, starting from naturally occurring IDP sequences.
The authors found that the net homotypic and heterotypic interactions must diﬀer between the optimized IDPs, as expected.
In many cases, these net interactions were found to depend primarily on the monomer frequencies, such that the immiscibility of the two phases was not affected by randomizing the sequences of the designed IDPs.
However, when the genetic algorithm was initialized using a particular naturally occurring IDP sequence in one of the coexisting phases, the patterning of the amino-acid residues in the optimized partner sequence was found to be crucial for achieving immiscibility.
The reasons for this dependence on sequence patterning in some, but not all, optimization scenarios are poorly understood.
Nonetheless, sequences generated via this approach could provide challenging test cases for the further development of analytical sequence-dependent theories.

\subsubsection{Patchy-polymer simulations}

``Patchy-polymer'' models, which encode one-to-one interactions between binding sites on specific monomers, are appropriate CG models for testing the predictions of associating fluid theory.
To explore the design rules underlying multiphasic systems with this class of models, \refcite{harmon2018differential} introduced a lattice-based CG model of poly-PRM and poly-SH3 multidomain proteins.
Proline-rich modules (PRMs) are short IDR sequence motifs that engage in specific interactions with folded SH3 domains~\cite{li2012phase}, and as such form one-to-one binding interactions.
Meanwhile, the linkers between motifs in the poly-PRM molecules and between the folded domains in the poly-SH3 molecules, respectively, were modeled either implicitly, representing ideal chains with Gaussian conformational statistics, or explicitly, using a variable number of lattice-site-occupying monomers.
Simulations were conducted using two types of poly-SH3 molecules, which competed for binding to the PRMs.
The authors found that differences in the linker properties, which tune the effective pairwise interactions between the molecules in the absence of the associative PRM/SH3 interactions, strongly affect the ability of the mixture to form immiscible condensed phases.
This observation is consistent with the finding of \refcite{pal2021subcompartmentalization} that excluded volume interactions are necessary for demixing.
By contrast, the interaction volume associated with the attractive PRM/SH3 interactions was found to play a less important role in determining the degree of immiscibility, in line with the predictions of associating fluid theory in the strong-binding limit ($\rho\Delta \gg 1$) of \eqref{eq:saft-chem-eq}.
\refcite{harmon2018differential} also showed that the interfaces of immiscible condensates are similarly affected by the linker properties, since molecules containing linkers with greater excluded volumes are preferentially driven towards interfaces with the dilute phase.

\subsubsection{Patchy-particle models}

``Patchy-particle'' models allow for simulations with a larger number of distinct molecular species, along with a greater diversity of associative interactions, due to their simplicity.
In complex mixtures with a variety of different associative interactions, it is useful to describe the collection of all possible one-to-one binding interactions by introducing an ``interaction network''~\cite{sanders2020competing}.
\refcite{espinosa2020liquid} explored this network concept using MD simulations of a 6-component mixture comprising 2, 3, and 4-valent patchy particles.
The authors considered a nearly fully connected network with almost all equivalent interaction strengths, leading to the formation of a single condensed phase in mixtures with equimolar parent concentrations.
Unsurprisingly, the density of associative bonding interactions in the condensed phase was found to correlate with the condensate stability, as measured by the critical temperature.
Simulations further revealed that high-valence molecules, which phase separate with high critical temperatures in single-component solutions, tend to increase the critical temperature of multicomponent condensates when added to mixtures of components with lower average valence.
Of direct experimental relevance, positive correlations were observed between the critical point of a molecular species in a single-component solution, its binding-site valence, and its partition coefficient with respect to a multicomponent condensate in a mixture with equimolar parent concentrations.

These observations can be understood qualitatively within the framework of associating fluid theory.
Making the simplification that all binding sites interact with one another via the same interaction volume, such that ${\Delta_{iA,jB} = \bar\Delta\,\forall i,A,j,B}$, the solution to \eqref{eq:saft-chem-eq} simplifies to $X_{iA} = \bar X \,\forall i,A$.
The associative contribution to the excess chemical potential (see \secref{sec:associating-fluid}) is thus ${\beta\mu_{\text{assoc},i} \approx m_i \log \bar X}$ in the condensed phase and negligible in the dilute phase, implying that the partition coefficient, \eqref{eq:PC}, is related to the binding-site valence by ${\text{PC}_i \propto \exp(m_i)}$.
An approximate relationship between the stability of the condensed phase and the average valence of the mixture follows by a similar argument.
The relatively small variations in interaction strengths in the simulated interaction network~\cite{espinosa2020liquid} can be considered as perturbations on these predictions.
However, variations in the geometric arrangements of the binding sites, and their relatively minor effects on the partition coefficients, are not captured at this level of theory.

The patchy-particle model of \refcite{espinosa2020liquid} was then extended to examine multiphase mixtures in \refcite{sanchez2021valency}.
The authors modified the interaction network by eliminating heterotypic associative interactions between select molecular species in order to construct immiscible condensates and multilayered structures.
Analogously to the results of \refcite{harmon2018differential}, simulations demonstrated that strong homotypic associative interactions lead to the formation of multiple immiscible condensates, while the introduction of strong heterotypic associative interactions tends to stabilize a single condensed phase.
However, mixtures with competing heterotypic interactions between weakly and strongly associating species showed evidence of multiphase condensate formation.

\section{Outlook and challenges}
\label{sec:outlook}

We have reviewed recent progress in the development of statistical and sequence-specific theories of multicomponent fluids and multiphase condensate formation.
Further advances in this area have the potential to reveal quantitative relationships between the molecular determinants of biomolecules, whether naturally occurring or engineered, and phase-separated self-organization in heterogeneous mixtures.
In particular, inverse-design strategies offer a promising approach for rationally and systematically identifying the physicochemical properties of biomolecular mixtures responsible for the assembly of complex---and biologically functional---condensates.
These theoretical and computational efforts will help to provide a roadmap for future experiments on heterogeneous biomolecular mixtures.

Nonetheless, many significant theoretical challenges remain to be explored, particularly with regard to the assumption of thermal equilibrium.
Future directions for theoretical and simulation advances in this field include:
\begin{enumerate}

\item \textit{Structuring and parameterizing multicomponent mixture models.}
Further development of statistical mixture models (see \secref{sec:mean-field}) will require incorporating information from sequence-dependent theories and simulations.
In this way, it will be possible to investigate the thermodynamic consequences of physically motivated and biomolecularly relevant correlations among interaction parameters in multicomponent fluids, as well as to move beyond pairwise mixture models.
Physicochemically motivated constraints should also be incorporated into inverse design approaches.

\item \textit{Extending sequence-dependent coarse-grained simulations and theories to multicomponent mixtures.}
Complementary insights can be gained by increasing the number of components in biomolecular mixtures treated using analytical theories or studied via coarse-grained molecular simulation (see \secref{sec:sequence-dependent}).
Simulations of recently developed CG IDP models~\cite{dignon2018sequence,dignon2018relation,latham2019maximum,joseph2021physics,tesei2021accurate,tesei2022improved,benayad2020simulation} have demonstrated impressive agreement with experiments on both single-chain and individual condensed-phase properties, suggesting that multicomponent simulations using these models may also be capable of predicting multiphase coexistence~\cite{chew2023thermodynamic} with similar accuracy.
Further improvement in the chemical accuracy of multicomponent simulations is likely to be achieved through multiscale approaches that incorporate all-atom simulations of ribonucleic condensates~\cite{zheng2020molecular,gruijs2022disease,galvanetto2022ultrafast}.
In future simulation studies, it will also be important to consider the role of competition between sequence-dependent clustering, aggregation, and LLPS behaviors, as observed in simple models of single-component heteropolymer solutions~\cite{statt2020model,rana2021phase,rekhi2023role}, in multicomponent mixtures.

\item \textit{Accounting for nonequilibrium effects due to kinetic barriers.}
Within the near-equilibrium framework, kinetic effects can lead to differences between the phase behavior that is observed in simulations and experiments and what is predicted at global thermodynamic equilibrium.
For example, nucleation pathways~\cite{shimobayashi2021nucleation} and slow rates of transitions between metastable states~\cite{garaizar2022aging,erkamp2022multiphase,lin2023dynamical} can affect the molecular compositions and multiphasic organization of phase-separated condensates on biologically relevant timescales.
The consequences of these nonequilibrium effects in systems with many components require further exploration.

\item \textit{Exploring differences in phase behavior at nonequilibrium steady states.}
Phase separation can also occur in fluids at nonequilibrium steady states (NESSs), which can arise due to chemostatted chemical reactions~\cite{soding2020mechanisms}.
Differences between thermal equilibrium and a NESS can manifest, for example, in the nucleation behavior~\cite{cho2023tuning,ziethen2022nucleation} as well as the growth and coarsening dynamics~\cite{zwicker2015suppression,zwicker2017growth,wurtz2018chemical,weber2019physics,kirschbaum2021controlling} of phase-separated droplets.
The implications of chemically driven NESSs for multiphase self-organization are largely unexplored.

\item \textit{Developing theoretical tools for emerging experimental applications.}
A variety of experimental platforms for manipulating biomolecular LLPS have recently been developed using ``designer'' peptides~\cite{simon2017programming,fisher2020tunable,heidenreich2020designer,baruch2023biomolecular}, nucleic acids~\cite{rovigatti2014accurate,hegde2023competition,leo2022pairing}, and non-biological polymers~\cite{lu2020multiphase}.
Chemically specific computational tools are needed to guide the rational design of multicomponent, multiphasic mixtures using these experimental platforms.
With a better understanding of condensate compositional control in heterogeneous environments, combined theoretical and experimental engineering approaches have the potential to bring about practical techniques for manipulating complex biological processes \textit{in vivo}~\cite{lyons2023functional}.
\end{enumerate}

In summary, LLPS can give rise to highly nontrivial spatial organization in multicomponent biomolecular fluids.
Nevertheless, considerable gaps persist in our understanding of the relationship between molecular-level properties and emergent phase behavior in heterogeneous mixtures.
Addressing this multifaceted question therefore represents an important way in which chemical theory and simulation can contribute to research at the forefront of molecular and cell biology, while helping to elucidate the origins of self-organization in living systems.

%% \begin{acknowledgments}
This work is supported by the National Science Foundation (DMR-2143670).
%% \end{acknowledgments}

\providecommand{\latin}[1]{#1}
\makeatletter
\providecommand{\doi}
  {\begingroup\let\do\@makeother\dospecials
  \catcode`\{=1 \catcode`\}=2 \doi@aux}
\providecommand{\doi@aux}[1]{\endgroup\texttt{#1}}
\makeatother
\providecommand*\mcitethebibliography{\thebibliography}
\csname @ifundefined\endcsname{endmcitethebibliography}
  {\let\endmcitethebibliography\endthebibliography}{}

\end{document}